\documentstyle[amsfonts,aps]{revtex}

\begin{document}
\large

\title{ Multi-channel phase-equivalent transformation and supersymmetry} 
\author{Andrey M. Shirokov}
\address{\normalsize \sl
Skobeltsyn Institute of Nuclear Physics, 
Moscow State University, 
Moscow, 119899 Russia\\
E-mail: shirokov@npi.msu.su}
\author{V. N. Sidorenko}
\address{\normalsize \sl Physics Department, 
Moscow State University, 
Moscow, 119899 Russia\\
E-mail: sidorenk@goa.bog.msu.su} 

\maketitle

\begin{abstract}
{\normalsize Phase-equivalent transformation of local
interaction is generalized to the multi-channel case. Generally, the
transformation does not change the number of the
bound states in the system and their energies. However, with a special
choice of the parameters, the transformation removes one of the bound
states and is equivalent to the
multi-channel supersymmetry transformation  recently suggested in
\cite{SpBaye}.  Using the transformation, it is also possible to add a
bound state to the discrete spectrum of the system at a given energy
$E<0$ if the angular momentum at least in one of the coupled channels
$l\ge 2$.}
\end{abstract}

\large
\section{Introduction}

Nucleon-nucleon, nucleon-cluster and cluster-cluster potentials are an
input for various microscopic calculations of nuclear structure and
reactions. Unfortunately, the exact form of the potentials describing
these interactions is unknown. It is conventionally supposed that the
interactions are local, that is, of course, an approximation
only. However, the available scattering data and bound states 
properties can be fitted with  approximately the same accuracy by
different local potentials. For example, there is a lot of so-called
realistic $NN$ potentials on the market describing
$NN$ scattering and deuteron properties with high accuracy. More, a
description of phenomenological data can be achieved with
the potentials very different in structure. In particular,
meson-exchange $NN$ potentials like the  Nijmegen one \cite{Nijmegen},
are known to have a short-range repulsive core in a triplet $s$ wave.
The same high-quality description of the nucleon-nucleon data is
provided by latest versions of 
Moscow potential \cite{MoscowProgr,Moscow3N} that does not have a
repulsive core but instead is deeply-attractive in the triplet $s$
wave at short distances and
supports an additional forbidden state. The possibility of alternative
description  of various cluster-cluster and nucleon-cluster interactions
by means of repulsive-core and
deeply-attractive potentials with forbidden states, is also well-known
(see, e.g., the discussion in \cite{MoscowProgr} and references
therein).

Principally it is possible
to distinguish experimentally between alternative
potentials studying their off-shell properties in interaction with an
additional particle.
The simplest probe is the photon, and as it was
shown in \cite{ppBremss,BremssSt.P.,BremssNPA}, the proton-proton
bremsstrahlung reaction $pp\to pp\gamma$ at the energy range of
350--400 MeV can be used to discriminate between various
nucleon-nucleon potentials. However  the $pp\to pp\gamma$ reaction has
not been examined experimentally in this energy range.

Another possibility is to study properties of three and four body
systems bound by two-body potentials of interest. From this point of
view, it looks like that we do not have at present satisfactory
nucleon-nucleon, cluster-nucleon and cluster-cluster potentials. It is
well-known that none of the realistic $NN$ potentials provides proper
binding of tritium or $^3$He. There are successful attempts in
generating phenomenological  three-nucleon interactions tuned to fit the properties
of light nuclei \cite{Argonne} (see also \cite{Argonne2} and
references therein). However, as it was shown in a detailed
study of  Picklesimer et al \cite{trinucl}, 
the effect of three-nucleon forces consistent with realistic
two-body ones on the binding energy of the triton is canceled by effects
of the virtual excitation of $\Delta $ isobars, etc. Hence the
trinucleon cannot be  satisfactorily described using known realistic two-body
potentials supplemented by three-body potentials consistent with
them. All calculations within three-body cluster models also fail to
reproduce the correct binding energy of three-cluster nuclear systems
with known local cluster-cluster and cluster-nucleon potentials
fitted to the corresponding scattering data.

To design a potential consistent  with two-body phenomenological data
and providing the correct binding of few-body systems, it seems
promising to make
use of phase-equivalent transformations depending on a
continuous parameter(s). Some attempts in this direction have been
performed using non-local phase-equivalent transformations.
The results of these attempts are encouraging: in Ref. \cite{IS} an
oversimplified $NN$ potential providing a satisfactory description of
$s$ wave $NN$ scattering date was fitted to reproduce exactly the
triton binding energy, while in Ref. \cite{6He} realistic
$n$--$\alpha$ potentials were tuned to reproduce various $^6$He
properties including the binding energy within the $\alpha+n+n$
cluster model.  The interactions suggested in Ref. \cite{IS,6He} are
non-local. Various applications (see, e.g., \cite{Garrido,Sof}) of
local phase-equivalent transformations to few-body problems were
restricted to the supersymmetry transformation
\cite{Andrianov,Sukumar,SUSY} that removes one of the bound states in
a two body system. The supersymmetry transformation
does not contain parameters and
cannot be used for fine tuning of the interaction of interest.

A local phase-equivalent transformation which preserves the number of
the bound states and depends on a continuous parameter, exists and is
well-known in the inverse scattering theory \cite{Newton}. 
Recently the effect of this transformation on the properties of three and
four nucleon systems was studied in Ref.~\cite{MSWVP} using as an
example a semi-realistic Malfliet-Tjon $NN$
interaction~\cite{MTIV}. It was shown in  Ref.~\cite{MSWVP} that a
slight phase-equivalent modification of $NN$ interaction is enough to
reproduce the trinucleon binding energy and to improve simultaneously the
description of four-nucleon binding. However the local
transformation was
developed for a single-channel case only and cannot be applied without
some approximations to realistic $NN$ interactions that mix triplet
$s$ and $d$ partial waves. Another drawback of the transformation is
that it involves a bound state
wave function and thus cannot be used to modify  $nn$ and $pp$
interactions and the $np$ interaction in all `non-deuteron' partial
waves.

Recently Sparenberg and Baye \cite{SpBaye} suggested a multi-channel
supersymmetry transformation. We use some ideas of Ref.  \cite{SpBaye}
to derive in what follows a multi-channel phase-equivalent
transformation which depends on continuous parameters. The
transformation can be treated as a generalization both of the
single-channel phase-equivalent 
transformation \cite{Newton} and of the multi-channel supersymmetry
transformation of Ref. \cite{SpBaye}. Generally, the
transformation does not change the number of the
bound states in the system and their energies. However, with a special
choice of the parameters, the transformation removes one of the bound
states and becomes equivalent to the
multi-channel supersymmetry transformation  suggested in
\cite{SpBaye}.  If the angular momenta in all coupled
channels are less than 2, a  parameter-dependent family of local interactions
phase-equivalent to the given initial one can be constructed by means
of the transformation even in the case when  the system does not have
a bound state. If the angular momentum at least in one of the coupled channels
$l\ge 2$, the transformation can be used to add a
bound state to the discrete spectrum of the system at a given energy
$E<0$. Having a bound state, one can construct a family of
phase-equivalent potentials and afterwards remove the bound state by the
supersymmetry version of the transformation. Thus, the suggested
transformation can be used in a multi-channel case to produce
phase-equivalent interactions without any restriction on the structure
of the discrete spectrum  of the system. In particular, the
transformation can be applied to the realistic $NN$ interaction in all
partial waves.

\section{General form of local multi-channel phase-equivalent transformation}


Multi-channel scattering and bound states we describe by 
Schr\"odinger equation
\begin{equation}
\sum_j \left( H_{ij}-E\delta_{ij} \right)\varphi_j(E,r) =0,
\label{Schr}
\end{equation}
where indexes $i$ and $j$ label channels, $E$ is the energy, the 
Hamiltonian
\begin{equation}
H_{ij}=\frac{\hbar^2}{2m}\left[-\frac{d^2}{dr^2} +
\frac{l_i(l_i+1)}{r^2}\right]\delta_{ij}+V_{ij}(r),
\label{Ham}
\end{equation}
$m$ is the reduced mass, and  $l_i$ stands for the angular momentum in
the channel $i$. We suppose that the potential
$V_{ij}(r)$  (i)~is Hermitian and (ii) at large 
distances it tends asymptotically to a diagonal constant matrix,
\begin{equation}
V_{ij}(r)\mathop{\longrightarrow}\limits_{r\to \infty}
\epsilon_i\:\delta_{ij}\,,
\label{Vlimit}
\end{equation}
where $\epsilon_i$ is a threshold energy in the channel $i$. We suppose 
that $\epsilon_1=0$ and $\epsilon_i\ge\epsilon_j$ if $i>j$.

Boundary conditions for the wave functions are
\begin{eqnarray}
&&\varphi_i(E,0)=0, \label{bound0}\\
&&\varphi_i(E,\infty)<\infty.\label{boundinfty}
\end{eqnarray}
Except for the discussion in section \ref{InSUSY}, we suppose that
there is at least one bound state in the system at the energy $E_0$. 
The corresponding wave function, $\varphi_i(E_0,r)$, is supposed to be 
normalized,
\begin{equation}
\sum_i\int\limits_0^\infty{\varphi_i}^*(E_0,s)\;\varphi_i(E_0,s)\;ds=1,
\label{Norm}
\end{equation}
where $^*$ denotes the complex conjugation. 
Of course, $\varphi_i(E_0,r)$ fits more severe
boundary condition  at $r\to\infty$  than (\ref{boundinfty}):
\begin{equation}
\varphi_i(E_0,\infty)=0.
\label{boundinfty-bound}
\end{equation}


We define the transformed potential $\tilde V_{ij}(r)$  as 
\begin{equation}
\tilde V_{ij}(r)=V_{ij}(r)+v_{ij}(r),
\label{tildeV}
\end{equation}
where 
\begin{equation}
v_{ij}(r)=-2C\,\frac{\hbar^2}{2m}\;\frac{d}{dr}\;
\frac{{\varphi_i}(E_0,r){\varphi_j}^*(E_0,r)}{A+C
\sum\limits_k\int\limits_a^r\left|{\varphi_k}(E_0,s)\right|^2\;ds}\:,
\label{vij}
\end{equation}
and $A$, $C$ and $a$ are arbitrary real parameters.

The main result of this paper can be formulated  as the following statement.

\begin{itemize}
\item {\sl The wave function 
\begin{equation}
\tilde{\varphi_i}(E,r)={\varphi_i}(E,r)
-C{\varphi_i}(E_0,r)\frac{
\sum\limits_k\int\limits_a^r{\varphi_k}^*(E_0,s)\:\varphi_k(E,s)\;ds}{A+C
\sum\limits_k\int\limits_a^r\left|{\varphi_k}(E_0,s)\right|^2\;ds}
\label{tildephi}
\end{equation} 
fits  {\bf inhomogeneous} multi-channel Schr\"odinger equation 
\begin{equation}
\sum\limits_j \left( \tilde H_{ij}- \right. 
\left.\vphantom{\tilde H}
E\delta_{ij} \right)\tilde\varphi_j(E,r) =
C\;\frac{\hbar^2}{2m}\;\frac{\varphi_i(E_0,r)}{A+C
\sum\limits_k\int\limits_a^r\left|{\varphi_k}(E_0,s)\right|^2\;ds}
\;{\cal W}(E_0,E;a),
\label{inhomo}
\end{equation}
where the Hamiltonian
\begin{equation}
\tilde H_{ij}=\:\delta_{ij}\:\frac{\hbar^2}{2m}\left[-\frac{d^2}{dr^2} +
\frac{l_i(l_i+1)}{r^2}\right]+\tilde V_{ij}(r)
\label{tildeHam}
\end{equation}
and the quasi-Wronskian}
\begin{equation}
{\cal W}(E_0,E;a)\equiv \sum_k\left[
{\varphi_k}^*(E_0,a)\;{\varphi_k}^{\prime}(E,a)-
{{\varphi_k}^*}^{\prime}(E_0,a)\;{\varphi_k}(E,a)\right].
\label{Wrons}
\end{equation}
\end{itemize}
We use prime to denote derivatives: $f'\equiv \frac{d}{dr}f$.

To prove the statement, one can verify Eq.~(\protect\ref{inhomo}) by
the direct calculation of\ \  
$\sum_j \left( \tilde H_{ij}-E\delta_{ij} \right)\tilde\varphi_j(E,r)$
using the definitions (\ref{tildeV})--(\ref{tildephi}), 
(\ref{tildeHam}) and (\ref{Wrons}) and other
formulas given above as well as the fact that the interaction
$V_{ij}(r)$ is Hermitian, $V_{ij}^*(r)=V_{ji}(r)$. 
The calculation is lengthy but straightforward. 

It is clear from (\ref{tildephi}) and (\ref{boundinfty-bound}) that
the suggested transformation is phase-equivalent at any energy $E>0$;
all the bound states supported by the initial potential $V_{ij}$ are
preserved by the transformation since the wave functions
$\tilde{\varphi_i}(E_b,r)$ for the corresponding energies $E_b<0$
(including $E_0$) fit both boundary conditions (\ref{bound0}) and
(\ref{boundinfty-bound}). However, the
denominator in the last term in (\ref{tildephi}) should be non-zero at
any distance $r$, and therefore  one should
be accurate in assigning values to arbitrary parameters
$A$, $C$ and $a$. This
requirement can be easily satisfied in a wide and continuous range of
parameter values.

\section{Particular cases of the phase-equivalent transformation}
\subsection{Homogeneous Schr\"odinger equation} 

Of course, we are mostly interested in phase-equivalent transformations that
results in homogeneous Schr\"odinger equation
\begin{equation}
\sum_j \left( \tilde H_{ij}-E\delta_{ij} \right)\tilde\varphi_j(E,r) =0
\label{homo}
\end{equation}
instead of the inhomogeneous Schr\"odinger equation (\ref{inhomo}). To
derive the transformation leading to Eq.~(\ref{homo}), we can fix the
parameters $A$, $C$, and $a$ in such a way that the r.h.s. of
Eq.~(\ref{inhomo}) will take zero value. The choice $C=0$ brings us to
the equivalent (contrary to phase-equivalent) transformation that is
of no interest.
Thus we should search for the
parameters that fit the equation 
\begin{equation}
{\cal W}(E_0,E;a)=0.
\label{Param-eq}
\end{equation}

Two obvious solutions of Eq.~(\ref{Param-eq}) are $a=0$ and
$a=\infty$. Various other solutions of 
Eq.~(\ref{Param-eq}) can be found for particular potentials
$V_{ij}(r)$. However, the non-zero finite solutions $a$
of  Eq.~(\ref{Param-eq}) are energy-dependent. With the solutions
$a(E)$ of  Eq.~(\ref{Param-eq}) we can obtain energy-dependent
potentials $\tilde V_{ij}(E;r)$ phase-equivalent to the initial
energy-independent potential $V_{ij}(r)$.  It may be interesting for some
applications, but we shall not discuss the energy-dependent
transformation and shall concentrate our attention on the solutions $a=0$ and
$a=\infty$.

The case $a=0$ presents a generalization of the single-channel
phase-equivalent transformation of Ref.~\cite{Newton}. For the bound
state at the energy $E_0$, the wave function obtained by means of the
transformation is of the form:
\begin{equation}
\tilde\varphi_i(E_0,r)=\frac{A\varphi_i(E_0,r)}{A+
C\sum\limits_j\int\limits_0^r\left|\varphi_j(E_0,s)\right|^2\;ds}\:.
\label{boundwf0}
\end{equation}
The wave function (\ref{boundwf0}) is not normalized. The
normalization constant can be easily calculated. The normalized bound
state wave function is
\begin{equation}
\sqrt{\frac{A+C}{A}}\;
\tilde\varphi_i(E_0,r)=\frac{\sqrt{A(A+C)}\;\varphi_i(E_0,r)}{A+
C\sum\limits_j\int\limits_0^r\left|\varphi_j(E_0,s)\right|^2\;ds}\:.
\label{boundwf0-N}
\end{equation}

It is interesting that the components of the bound state wave function in all
channels are modified by the transformation sinchronically: all the
components $\varphi_i(E_0,r)$ are multiplied by the same multiplier
$\vphantom{\displaystyle\int}
{\sqrt{A(A+C)}} \Biggl( {A+  
C\sum\limits_j\int\limits_0^r\left|\varphi_j(E_0,s)\right|^2\;ds}
\Biggr)^{-1}$.
Nevertheless the relative weight of the components  $\varphi_i(E_0,r)$
in the norm of the total multi-channel wave function can be changed by the
transformation. 

Now let us discuss the case $a=\infty$. The transformed wave function in this
case is of the form:
\begin{equation}
\tilde\varphi_i(E,r)=\varphi_i(E,r)\; - \;
\frac{C\varphi_i(E_0,r)\sum\limits_j\int\limits_\infty^r
{\varphi_j}^*(E_0,s)\;\varphi_j(E,s)\;ds}{A+
C\sum\limits_j\int\limits_\infty^r\left|\varphi_j(E_0,s)\right|^2\;ds}\:.
\label{wfoo}
\end{equation}
If $E\ne E_0$, the functions $\varphi_i(E,r)$ and $\varphi_i(E_0,r)$
are orthogonal:
\begin{equation}
\int\limits_0^\infty{\varphi_i}^*(E_0,s)\;\varphi_i(E,s)\;ds=0.
\label{Orth}
\end{equation}
With the help of (\ref{Orth}) and (\ref{Norm}), we can rewrite
(\ref{wfoo}) as
\begin{equation}
\tilde\varphi_i(E,r)=\varphi_i(E,r)\; - \;
\frac{C\varphi_i(E_0,r)\sum\limits_j\int\limits_0^r
{\varphi_j}^*(E_0,s)\;\varphi_j(E,s)\;ds}{A-C+
C\sum\limits_j\int\limits_0^r\left|\varphi_j(E_0,s)\right|^2\;ds}\:.
\label{wfoo-2}
\end{equation}
It is seen from (\ref{wfoo-2}) that the case $a=\infty$ is identical
(up to the redefinition of the parameter $A\to A+C$)
to the case $a=0$ if 
$E\ne E_0$. It is clear, however, 
that after the redefinition of the parameter  $A\to A+C$, the potential
$v_{ij}(r)$ obtained with $a=\infty$ becomes identical to the potential
$v_{ij}(r)$ corresponding to the case $a=0$. Hence the the case
$a=\infty$ appears to be  identical to the case $a=0$ at any energy
$E$ including $E=E_0$. To demonstrate this explicitly, let us examine
the wave function $\tilde\varphi_i(E_0,r)$ in the case
$a=\infty$. Substituting $E$ by $E_0$ in (\ref{wfoo}) we obtain:
\begin{eqnarray}
\tilde\varphi_i(E_0,r)&=&\frac{A\varphi_i(E_0,r)}{A+
C\sum\limits_j\int\limits_\infty^r\left|\varphi_j(E_0,s)\right|^2\;ds}
                 \label{boundwfoo}     \\
&=&\frac{A\varphi_i(E_0,r) \vphantom{\displaystyle\int} }{A-C+
C\sum\limits_j\int\limits_0^r\left|\varphi_j(E_0,s)\right|^2\;ds}\; .
\label{boundwfoo-1}
\end{eqnarray}
Replacing $A$ by $A+C$ and normalizing the wave function
(\ref{boundwfoo-1}), we obtain the expression  (\ref{boundwf0-N}).

\subsection{Supersymmetry}
Let us discuss a particular choice of parameters: $C=1$, $a=\infty$,
and $A=1$. The wave function in this case is  
\begin{eqnarray}
\tilde\varphi_i(E,r)
&=&\varphi_i(E,r)+
\frac{\varphi_i(E_0,r)\sum\limits_j\int\limits_r^\infty
{\varphi_j}^*(E_0,s)\;\varphi_j(E,s)\;ds}{
\sum\limits_j\int\limits_0^r\left|\varphi_j(E_0,s)\right|^2\;ds}
\label{SUSY-Baye}\\
&=&\varphi_i(E,r)-
\frac{   \vphantom{\displaystyle\int\limits^0}
\varphi_i(E_0,r)\sum\limits_j\int\limits_0^r
{\varphi_j}^*(E_0,s)\;\varphi_j(E,s)\;ds}{
\sum\limits_j\int\limits_0^r\left|\varphi_j(E_0,s)\right|^2\;ds}\:.
\label{SUSY}
\end{eqnarray}
Equation~(\ref{SUSY-Baye}) can be used at any energy $E$
while Eq.~(\ref{SUSY}) is applicable only if $E\ne E_0$.
In the case $E=E_0$, the wave function can be rewritten in a simpler form as 
\begin{equation}
\tilde\varphi_i(E_0,r)=
\frac{\varphi_i(E_0,r)}{
\sum\limits_j\int\limits_0^r\left|\varphi_j(E_0,s)\right|^2\;ds}\; .
\label{bound-SUSY}
\end{equation}

Equation (\ref{SUSY-Baye}) is just the Eq.~(4) of
Ref.~\cite{SpBaye}.  In Ref.~\cite{SpBaye}, 
Sparenberg and Baye suggested a multi-channel supersymmetry
transformation. Thus equations (\ref{SUSY})--(\ref{bound-SUSY})
describe the multi-channel supersymmetry
transformation, or, in other words, the multi-channel supersymmetry
transformation is a particular case of the phase-equivalent
multi-channel transformation discussed in this paper that corresponds
to the particular choice of the parameters. Let us discuss how it works.

It is clear from Eq.~(\ref{bound-SUSY}) that 
$\left|\tilde\varphi_i(E_0,r)\right|\to\infty$ as $r\to 0$. Hence at the
energy $E_0$, the wave function $\tilde\varphi_i(E_0,r)$ does not match
the required boundary condition (\ref{bound0}) at $r=0$. At the same time,
$\tilde\varphi_i(E_0,r)$ fits the boundary condition
(\ref{boundinfty-bound}) at $r=\infty$. Therefore it is impossible to
construct another solution of the Schr\"odinger equation (\ref{homo})
consistent with both  boundary conditions at the energy
$E=E_0$.  As a result the
phase-equivalent transformation removes the bound state at $E=E_0$. At
the same time, it is clear from (\ref{SUSY}) that for all energies 
$E\ne E_0$, zero in the denominator arising in the limit $r\to 0$ is
canceled by the zero in the 
numerator and the wave function  (\ref{SUSY}) matches the boundary
conditions at the origin and at the infinity both at once. So, the
transformation in this case removes the bound 
state at $E=E_0$ but not any of the other bound states, while 
the $S$-matrix at any energy $E>0$ is unchanged.

Of course, the supersymmetry transformation can be also formulated in
the case $a=0$. It is interesting that the bound state in this case is
removed by a different mechanism. Suppose $A=0$ and $C$ is arbitrary. The wave
function at any energy $E$ in this case may be written as 
\begin{equation}
\tilde\varphi_i(E,r)=\varphi_i(E,r)-
\frac{\varphi_i(E_0,r)\sum\limits_j\int\limits_0^r
{\varphi_j}^*(E_0,s)\;\varphi_j(E,s)\;ds}{
\sum\limits_j\int\limits_0^r\left|\varphi_j(E_0,s)\right|^2\;ds}\:.
\label{SUSY-a-0}
\end{equation}
It is seen that at $E=E_0$ the wave function
$\tilde\varphi_i(E_0,r)\equiv 0$. 

We used the  boundary condition (\ref{bound0}) to construct the
supersymmetry transformation: the bound state is removed because
for some particular parameter values the 
wave function $\tilde\varphi_i(E_0,r)$ diverges at the
origin and appears to be inconsistent with (\ref{bound0}).
One can suppose that it is  also possible to use the boundary condition at
$r=\infty$ to remove the bound state and to construct another
supersymmetry transformation. It is not so. Let us discuss the case of
$a=\infty$, $A=0$ and arbitrary $C$. As is seen from (\ref{wfoo}),  
$\tilde\varphi_i(E_0,r)\equiv 0$ in this
case, thus the bound state is removed. However the
transformation is no more phase-equivalent. Really, at
energies $E> E_0$ the
last term in  (\ref{wfoo}) does not vanish at $r\to\infty$ 
and provides an additional phase shift, or, in other words, it modifies
the $S$-matrix.

An example of an application of the multi-channel supersymmetry
transformation to the Moscow $NN$ potential can be found in Ref.~\cite{SpBaye}.

\subsection{Inverse supersymmetry\label{InSUSY}}

A transformation that adds a bound state to the discrete spectrum of
the system and leaves unchanged the $S$-matrix and the energies of all
bound states supported by the initial Hamiltonian, we shall refer to
as  inverse supersymmetry transformation. 

Let us suppose that there is no bound state at the energy $E_0<0$. By 
$\varphi_i(E_0,r)$ we now denote the wave function at energy $E_0$
that matches the boundary condition (\ref{boundinfty-bound}) at
infinity but diverges at the origin as $r^{-l_i}$ 
(see, e.g.\footnote{The $r^{-l}$ divergence of the
wave functions at the origin is derived in Ref.~\protect\cite{Landavshits}
for the single-channel case only. However  the derivation of the
$r^{-l}$ rule of Ref.~\protect\cite{Landavshits} can be easily generalized to
the multi-channel case,  at least for the potentials that
do not diverge in the origin.}, \cite{Landavshits})
where $l_i$ is the angular momentum in the channel $i$. 

Having $\varphi_i(E_0,r)$, we can use our transformation  to obtain the
homogeneous Schr\"odinger equation (\ref{homo}) in the case
$a=\infty$. The transformed wave function $\tilde\varphi_i(E,r)$ is
given by (\ref{wfoo}). It is seen from  (\ref{wfoo}) that
$\tilde\varphi_i(E,r)$ does not diverge in the origin and matches the
boundary conditions both at the origin and at infinity, at any energy
$E\ne E_0$. For $E=E_0$, the transformed wave function
$\tilde\varphi_i(E_0,r)$  is given by (\ref{boundwfoo}). It is clear that
$\tilde\varphi_i(E_0,r)$ at the origin is proportional to
$r^{2L-l_i-1}$ where $L=\max\{l_i\}$. Hence, $\tilde\varphi_i(E_0,r)$
matches the boundary condition (\ref{bound0}) if $L\geq 2$ and is not
consistent with  (\ref{bound0}) if $L\leq 1$. Therefore our
transformation with  $\varphi_i(E_0,r)$ irregular at the origin, is the
inverse supersymmetry transformation in the case $L\geq 2$. In
the case $L\leq 1$ the transformation appears to be a phase-equivalent
transformation that does not make use of the bound state and can be
applied to the system that does not support a bound state. If the
transformation is applied to the free Hamiltonian with
$V_{ij}(r)\equiv 0$ in the $s$ or $p$
partial wave, it produces a non-zero `transparent' potential $\tilde
V_{ij}(r)$ that provides phase shift $\delta=0$ at any energy
$E$. The multi-channel version of the transformation couples $s$
and $p$ partial waves to produce a two-channel `transparent'
interaction that provides the $S$-matrix of the form $S_{ij}=\delta_{ij}$.

It is interesting that the inverse supersymmetry transformation is not
unique: we have three parameters $E_0$,
$A$ and $C$ that provide a family of inverse supersymmetry
partner potentials. Contrary to it, the supersymmetry transformation is unique;
however, it can be used in combination with the 
phase-equivalent transformation to construct a family of
potentials phase-equivalent to the initial one but not supporting one
of the bound state.

The multi-channel inverse supersymmetry transformation is discussed
in more detail  
in a very recent paper of  Leeb et al~\cite{LSoSpB}. One can find in
this paper examples of the applications of the transformation to
realistic $NN$ potentials. This
transformation is discussed in Ref.~\cite{ZhakhCh}, too; in
particular the authors of  Ref.~\cite{ZhakhCh} also conclude
that it is possible to create a new bound state by means of the
phase-equivalent transformation only in the case  $L\ge 2$.

\section{Conclusions}

We derived a multi-channel phase-equivalent transformation that
can be used without restrictions on the structure of the discrete
spectrum of the system in various scattering problems like $NN$
scattering, nucleon-cluster or cluster-cluster
scattering. The multi-channel supersymmetry and inverse supersymmetry
transformations appear to be particular cases of the suggested general
phase-equivalent transformation corresponding to particular choices of
the parameter values. The inverse supersymmetry transformation is
possible if only the orbital angular momentum $l_i\ge 2$ at least in one of the
coupled channels. It is interesting to note that from the
point of view of the $NN$ system, this means that a deep attractive
$NN$ potential supporting an additional forbidden state like Moscow
$NN$ potential, can be constructed by the inverse supersymmetry
transformation of the realistic meson-exchange potential with
repulsive core only due to the $d$ wave admixture in the deuteron wave
function.    

With the help of the suggested transformation, one can construct a
family of phase-equivalent potentials depending on continuous
parameters. Such families may be very useful for fine tuning of the
interaction aimed to fit not only two-body observables but also three-
and few-body ones. If the system has at least one bound state, the
phase-equivalent potential family is constructed using directly formulas 
(\ref{tildeV}) and (\ref{vij}). One can construct phase-equivalent
single- or multi-channel potential families also in the case when
there are no bound states in the system: if all channel orbital angular momenta
$l_i\leq 1$, one can apply directly the transformation with the irregular
function $\varphi_i(E_0,r)$; if at least one of the channel orbital
angular momenta 
$l_i\geq 2$, one can produce a bound state using inverse supersymmetry 
at the first stage and remove the bound state at the last stage with
the help of supersymmetry version of the transformation. So, one can,
for example, 
construct a family of phase-equivalent potentials for any combination
of coupled partial waves in the $NN$ system. 

	We hope that the suggested transformation will be useful in
various few-body applications.\\

{\bf Acknowledgements}. We are thankful to A. I. Mazur, A. Mondragon,
V.~N.~Pomerantsev, D.~L.~Pursey, Yu. F. Smirnov, J. P. Vary, 
T.~A.~Weber, and  B.~N.~Zakhar'ev  for stimulating discussions. The
work was supported in part by the
State Program ``Universities of Russia'', project No~992306 and  by 
the Competitive Center at St.~Petersburg State University.


\begin{thebibliography}{99}

\bibitem{Nijmegen}  V. G. J. Stoks, R. A. M. Klomp, C. P. F. Terheggen, and
J. J. de Swart, 
{\sl Phys. Rev. }{\bf C 49}, 2950 (1994).

\bibitem{MoscowProgr}  V.~I.~Kukulin and V.~N.~Pomerantsev, 
{\sl Progr. Theor. Phys. }{\bf 88}, 159 (1992).

\bibitem{Moscow3N}  V. I. Kukulin, V. N. Pomerantsev, A. Faessler, A. J.
Buchmann, and E.~M.~Tursunov, 
{\sl Phys. Rev.} {\bf C57}, 535 (1998).

\bibitem{ppBremss}  V. G. Neudatchin, N. A. Khokhlov, A.~M.~Shirokov, and   
V.~A.~Knyr, 
{\sl Yad. Fiz. }{\bf 60}, 1086
(1997) [{\sl Phys. At. Nucl. }{\bf 60}, 971 (1997)].

\bibitem{BremssSt.P.}  A.~M.~Shirokov, 
{\sl in: Proc.  XIth Int. Workshop on Quantum Field Theory and High
Energy Phys.} (ed. B.~B.~Levtchenko), p.~397 (Moscow, 1997).

\bibitem{BremssNPA}  N. A. Khokhlov, V. A. Knyr, V. G. Neudatchin, and A. M.
Shirokov, 
{\sl Nucl. Phys.} {\bf A~629}, 218 (1998);
%
{\sl Phys. Rev.} {\bf C~62}, 054003 (2000). 

\bibitem{Argonne}  B. S. Pudliner, V. R. Pandharipande, J. Carlson,
S. C. Pieper, and R.~B.~Wiringa, {\sl Phys. Rev.} {\bf C 56},
1720 (1997).  

\bibitem{Argonne2} R. B. Wiringa, {\sl Nucl. Phys.} {\bf A 631}, 70c
(1998) 

\bibitem{trinucl}  A.~Picklesimer, R.~A.~Rice, and R.~Brandenburg.
{\sl Phys. Rev. Lett. }{\bf 68}, 1484 (1992);
{\sl Phys. Rev. }{\bf C 45}, 547 (1992);
2045 (1992);
2624 (1992);
{\bf 46}, 1178 (1992).

\bibitem{IS} A. M. Shirokov, Yu. F. Smirnov, and S. A. Zaytsev, 
{\sl Revista Mex. Fis.} {\bf 40}, Supl. {\bf 1}, 74 (1994).

\bibitem{6He} Yu. A. Lurie and A. M. Shirokov, {\sl Izv. RAN, Ser. Fiz.}
{\bf 61}, 2121 (1997) [{\sl  Bul. Rus. 
Acad. Sci., Phys. Ser.} {\bf 61}, 1665 (1997)].

\bibitem{Garrido} E. Garrido, D. V. Fedorov, and A. S. Jensen, {\sl
Nucl. Phys.} {\bf A~650}, 247 (1999).

\bibitem{Sof} H. Fiedeldey, S. A. Sofianos, A. Papastylianos, K. Amos,
and L. J. Allen, {\sl Phys. Rev. } {\bf C 42}, 411 (1990).

\bibitem{Andrianov} A. A. Andrianov, N. V. Borisov, and M. V. Ioffe,
{\sl Phys. Lett.} {\bf A 105}, 19 (1984).

\bibitem{Sukumar} C. V. Sukumar, {\sl J. Phys. } {\bf A 18}, 2937
(1985).

\bibitem{SUSY} D. Baye, {\sl Phys. Rev. Lett.} {\bf 58}, 2738 (1987).

\bibitem{Newton} R. G. Newton, {\em Scattering theory of waves and 
particles,  2nd. ed.} (Springer-Verlag, New York, 1982).

\bibitem{MSWVP} A.~I.~Mazur, A.~M.~Shirokov, T.~A.~Weber, J.~P.~Vary,
and D.~L.~Pursey,
{\sl in: Proc. XIV Int. Workshop on High Energy
Physics and Quantum Field Theory (QFTHEP'99)} 
(eds. B.B.Levchenko and  V.I.Savrin), p.~531 (Moscow, MSU-Press, 2000).

\bibitem{MTIV}  R. A. Malfliet and T. A. Tjon, {\sl Nucl. Phys.}  {\bf A~127},
161 (1969); {\sl Ann. Phys. (N.Y.)} {\bf 61}, 425 (1970).

\bibitem{SpBaye} J. M. Sparenberg and D. Baye, {\sl Phys. Rev. Lett.} 
{\bf 79}, 3802 (1997).

\bibitem{Landavshits} L. D. Landau and E. M. Lifshitz, {\em Quantum
mechanics: non-relativistic theory} (Pergamon Press, New York, 1977).

\bibitem{LSoSpB} H. Leeb, S. A. Sofianos, J.-M. Sparenberg, D. Baye,
{\sl nucl-th/0008054} (2000).

\bibitem{ZhakhCh} B.~N.~Zakhar'ev and V.~M.~Chabanov, {\sl
Phys. Part. Nucl.} {\bf 30}, 111 (1999). 

\end{thebibliography}
\end{document}